\documentclass[onecolumn]{revtex4}
\usepackage{amsmath,amssymb,graphics,epsfig,subfigure}
\usepackage{color}
\usepackage[colorlinks,linkcolor=red,anchorcolor=red,citecolor=green]{hyperref}
\usepackage{setspace}
\usepackage{booktabs}
\usepackage{amsfonts}
\usepackage{bbm}
\usepackage{graphicx}
\usepackage{pgfplots}
\usepackage{mathrsfs}
\usepackage{color}
\usepackage{verbatim}
\usepackage{cancel}
\usepackage{float}
\usepackage{xcolor}

\begin{document}
	\thispagestyle{empty}
\begin{center}
	\title{Complexification of the Kerr-AdS black hole thermodynamic phase transition}
	
	\author{Fei Guo$^{1}$, Zhen-Ming Xu$^{1,2,3,4,}$\footnote{zmxu@nwu.edu.cn}, and Robert B. Mann$^{5,}$\footnote{rbmann@uwaterloo.ca}
	\vspace{6pt}\\}

\affiliation{$^{1}$School of Physics, Northwest University, Xi'an 710127, China\\
	$^{2}$Peng Huanwu Center for Fundamental Theory, Xi'an 710127, China\\
	$^{3}$Shaanxi Key Laboratory for Theoretical Physics Frontiers, Xi'an 710127, China\\
	$^{4}$Fundamental Discipline Research Center for Quantum Science and technology of Shaanxi Province, China\\
    $^{5}$Department of Physics and Astronomy, University of Waterloo, Waterloo, Ont. Canada N2L 3G1}
	
	\begin{abstract}
In this paper, we establish a systematic complex-analytical framework for investigating the thermodynamics of high-dimensional singly spinning Kerr-AdS black holes. By performing an analytical continuation of the black hole horizon radius into the complex plane, we map the thermodynamic behavior for the black hole to the zeros distribution of the corresponding analytical function. In the region below the critical point, this approach provides the criterion for identifying phase transitions and reveals that the zeroth-order phase transition does not alter the underlying thermodynamic topological structure of the black hole. In the supercritical region, based on the Lee-Yang theory of phase transition, we construct the Widom line to characterize the supercritical crossover behavior for the Kerr-AdS black holes. This study provides a new perspective on the thermodynamic analysis of black holes.
	\end{abstract}

\maketitle
\end{center}

\section{Introduction}
The proposal of the Hawking radiation and the subsequent formulation of Bekenstein-Hawking entropy established the foundation of black hole thermodynamics~\cite{Bekenstein1973,Bardeen1973,Hawking1983}. In anti-de Sitter (AdS) spacetime, black hole systems exhibit various phase transitions; the most prominent example is the Hawking-Page phase transition, which is interpreted as the confinement/deconfinement phase transition of the dual quark-gluon plasma~\cite{Witten1998,Maldacena:1997re}. The formulation of the extended phase space, where the (negative) cosmological constant is treated as (positive) thermodynamic pressure and its conjugate variable as thermodynamic volume, has revealed a rich landscape of thermodynamic phase behavior in black holes~\cite{Kastor2009,Mann2025,Mann:2026nqs,Kubiznak:2016qmn,Kubiznak:2012wp,Dolan2011}. For charged and rotating black holes in dimensions $d < 6$, the system exhibits a first-order small-large black hole phase transition, which is formally analogous to the liquid-gas phase transition of a van der Waals fluid. Moreover, reentrant phase transitions (a phenomenon typically associated with multicomponent fluids) have been identified in four-dimensional Born-Infeld black holes as well as in rotating black holes for $d \ge 6$~\cite{NARAYANAN:1994,Gunasekaran:2012dq,Monteiro2009,Altamirano:2013ane,Frassino:2014pha,Altamirano:2014tva,Wei2016}.

Conventional studies of phase transitions are confined to the real parameter space, in which the equation of state in the real domain---defined by physical temperature, pressure, and horizon radius--- is solved, and real-valued Gibbs free energy profiles are computed to assess thermodynamic features. Although these approaches delineate macroscopic phase boundaries, they are insufficient for resolving analytic properties near the critical point due to the singularities of thermodynamic functions (or their derivatives).

A novel framework incorporating complex analysis into black hole thermodynamics has been developed to provide a unified description of black hole phase transition behaviors. Our main considerations are twofold. (i) On the one hand, the recent proposals~\cite{Xu2024,Wang:2023qxw,Barzili2024} of the black hole thermodynamic free energy landscape and the generalized free energy scheme map the analysis of black hole thermodynamic phase transitions into an extremum distribution problem of a real-valued potential function, namely, the zeros of the derivative of the potential function. (ii) On the other hand, the renowned Lee-Yang theory of phase transitions in statistical physics provides a unified analytical framework for thermodynamic phase transitions, the essence of which is the distribution of zeros  of the thermodynamic partition function~\cite{Xu:2025jrk}. These two lines of inquiry ultimately reduce to the study of zeros of functions, and thus it is natural to extend the analysis to the complex domain, with the expectation that the zero structures of analytic functions may reflect the analytic properties of black hole thermodynamic phase transitions.

In the region below the critical point, this approach constructs a complex free energy via the generalized Maxwell equal-area law by employing analytic continuation and proposes the ``Phase transition/Riemann surface'' correspondence~\cite{Xu2024}, thereby elucidating the universal characteristics of black hole thermodynamics. In the supercritical region, a recent study~\cite{Xu:2025jrk} took the projection of complex Lee-Yang zeros on the real phase plane to construct the Widom line structure, which marks the crossover line of the (supercritical) phase transition between large-like and small-like black holes. This approach has been successfully extended to more black hole systems~\cite{Li2026,Guo2026,Ou2026,Awal2026,Lee2026}.
	
Motivated by above complex analysis, in this paper, we extend this approach to the Kerr-AdS black hole.
We investigate Kerr-AdS black holes individually within the framework of complex analysis in 4, 5, and 6 spacetime dimensions. By characterizing their analytic behavior in both the subcritical and supercritical regions, we aim to unveil a more comprehensive picture of the universal thermodynamic properties inherent to these black hole systems. In what follows, we first introduce the complex-analytic approach for the subcritical region, and then utilize the approach described above  to identify phase transition behaviors. Subsequently, based on the Lee-Yang theory, we analyze the distribution of Lee-Yang zeros in the complex plane and construct the corresponding Widom structure. Finally, we summarize our primary results and discuss potential avenues for future investigation.

\section{Thermodynamics and phase structure}
\label{sec:thermo_found}
	
We begin by presenting the thermodynamic foundations and phase structure of $d$-dimensional singly spinning Kerr-AdS black holes.	The metric is given by~\cite{Altamirano:2013ane,Gibbons2004,Caldarelli2000}
\begin{equation}
ds^2 = -\frac{\Delta}{\rho^2} \left( dt - \frac{a}{\Xi} \sin^2\theta d\varphi \right)^2 + \frac{\rho^2}{\Delta} dr^2 + \frac{\rho^2}{\Sigma} d\theta^2 + \frac{\Sigma \sin^2\theta}{\rho^2} \left[ a dt - \frac{(r^2 + a^2)}{\Xi} d\varphi \right]^2 + r^2 \cos^2\theta d\Omega_{d-2}^2,
\end{equation}
where $d\Omega_{d-2}^2$ denotes the metric of a unit $(d-2)$-dimensional sphere and
\begin{align}
\begin{aligned}
\Delta &= (r^2 + a^2)\left(1 + \frac{r^2}{l^2}\right) - 2mr^{5-d},\\
\Sigma &= 1 - \frac{a^2}{l^2} \cos^2\theta,\\
\Xi &= 1 - \frac{a^2}{l^2}, \\
\rho^2 &= r^2 + a^2 \cos^2\theta,
\end{aligned}
\end{align}
here $l$ and $a$ denote the AdS radius and the rotation parameter, respectively. In the extended phase space, the cosmological constant is treated as an independent thermodynamic pressure, with the relationship
\begin{equation}
P = -\frac{\Lambda}{8\pi} = \frac{(d-1)(d-2)}{16\pi l^2}.
\end{equation}
The first law of thermodynamics and the Smarr relation can be expressed in terms of the event horizon $r_h$ which is the largest root of the metric function $f(r) = 0$,	
\begin{equation}
dM = T dS + \Omega_H dJ + V dP,
\end{equation}
\begin{equation}
\frac{d-3}{d-2} M = T S + \Omega_H J - \frac{2}{d-2} VP,
\end{equation}
where the various thermodynamic quantities are
\begin{align}
M &= \frac{\omega_{d-2}}{4\pi} \frac{m}{\Xi^2} \left( 1 + \frac{(d-4)\Xi}{2} \right), \\
V &= \frac{r_h A}{d-1}[1 + \frac{a^2}{\Xi} \frac{1 + {r_h^2}/{l^2}}{(d-2)r_h^2}],\\
J &= \frac{\omega_{d-2}}{4\pi} \frac{ma}{\Xi^2},\label{eq:J} \\
\Omega_H &= \frac{a}{l^2} \frac{r_h^2 + l^2}{r_h^2 + a^2}, \\
T &= \frac{1}{2\pi} \left[ r_h \left( \frac{r_h^2}{l^2} + 1 \right) \left( \frac{1}{a^2 + r_h^2} + \frac{d-3}{2r_h^2} \right) - \frac{1}{r_h} \right], \\
S &= \frac{\omega_{d-2}}{4} \frac{(a^2 + r_h^2) r_h^{d-4}}{\Xi} = \frac{A}{4},
\end{align}
and $\omega_d = 2\pi^{(d+1)/2}/\Gamma\left((d+1)/2\right)$ represents the volume of a unit $d$-dimensional sphere~\cite{Cvetic2011} .
	
According to the standard thermodynamic definition, the Gibbs free energy can be obtained as
\begin{equation}
G=M-T S=\frac{\omega_{d-2} r_h^{d-5}}{16\pi \Xi^2}\left( 3a^2 + r_h^2-\frac{(r_h^2 - a^2)^2}{l^2}+\frac{3a^2 r_h^4 + a^4 r_h^2}{l^4}\right).
\end{equation}
		
\begin{figure}[htbp]
		\centering
		\includegraphics[height=50mm]{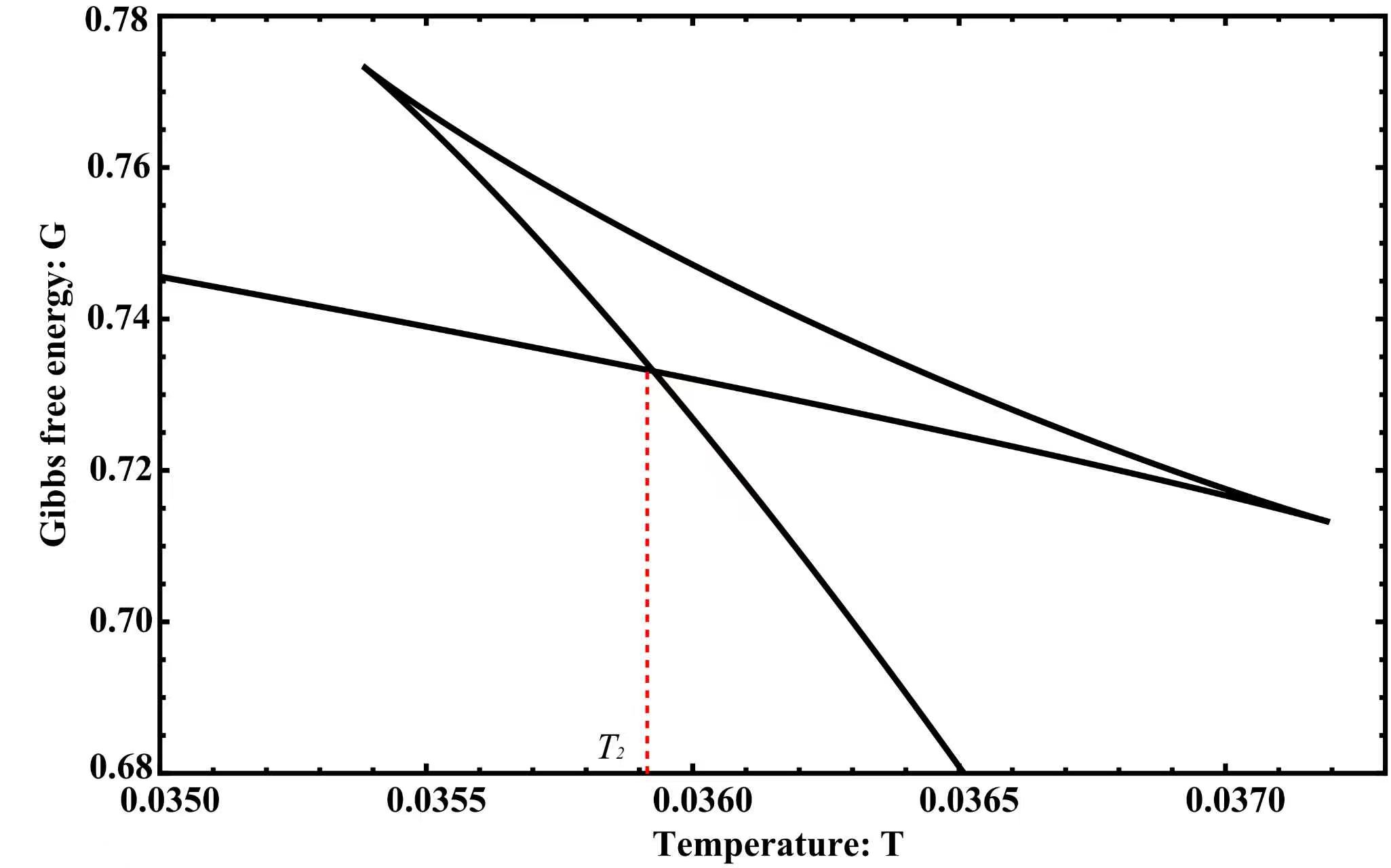}\hfill
		\includegraphics[height=50mm]{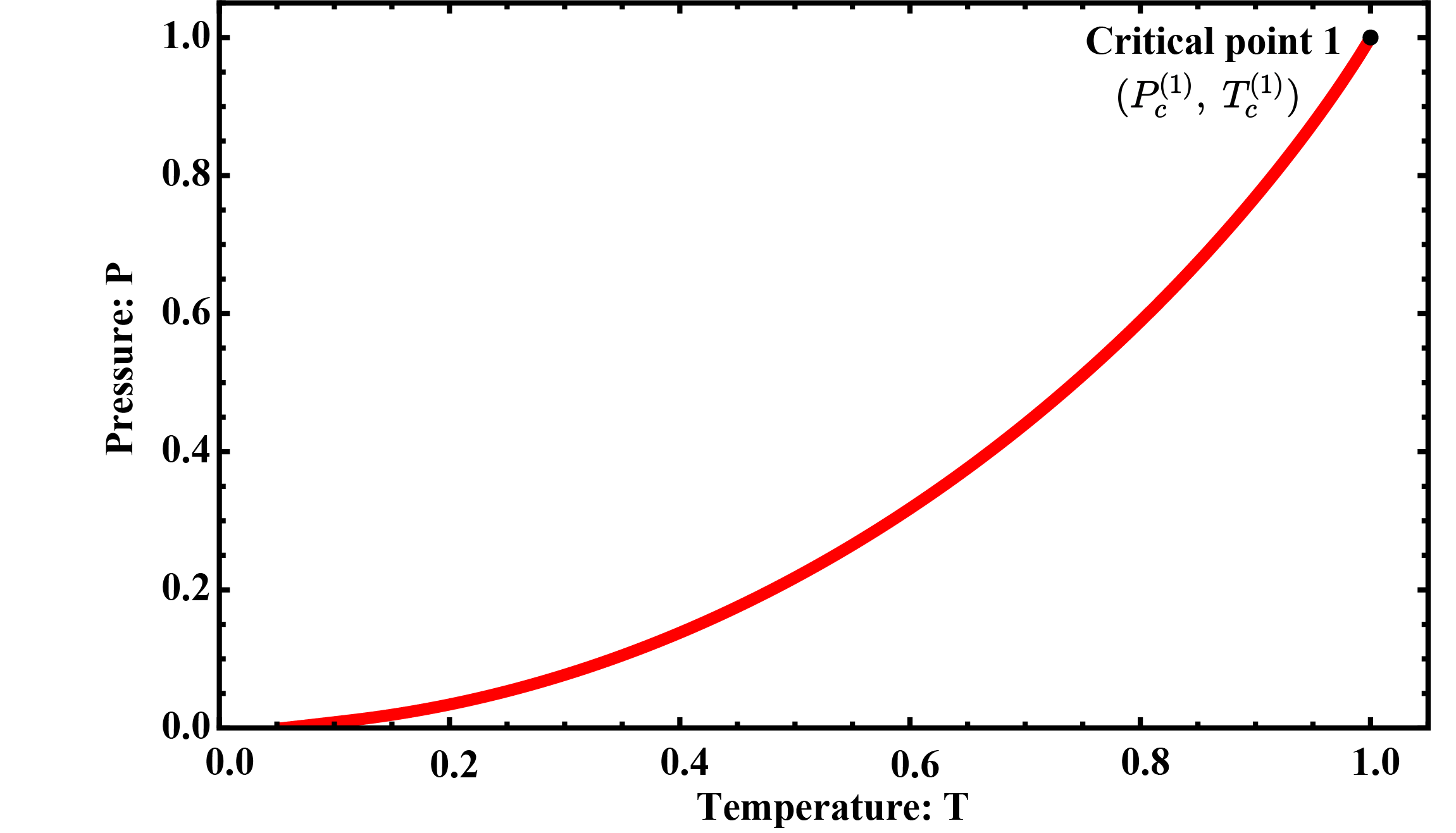}
		\caption{Phase diagram of Kerr-AdS black hole with $d=4$ and in the left diagram, we set $J=1$ and the pressure $P=0.002$.}
		\label{fig1_aligned}
	\end{figure}

The Gibbs free energy depends on the external parameter $J$. We can plot it for fixed $J$ parametrically (to simiplify the subsequent analysis, we set $J=1$). Specifically, $a$ is expressed as $a(J, r_h, P)$ via Eq.~(\ref{eq:J}), and substituting this well-behaved solution into the expressions for $G$ and $T$ allows us to represent both thermodynamic quantities as functions of $P$ and $r_h$~\cite{Altamirano:2013ane}. The behavior of $G$ depends significantly on the dimension $d$. For $4d$ and $5d$, the situations are same and we take the $4d$ case as a representative, shown in Fig.~\ref{fig1_aligned}. It exhibits standard swallowtail behavior when the temperature decreases to $T_2$, the black hole undergoes a phase transition from small black hole to large black hole, displaying the typical first-order phase transition behavior analogous to that of $4d$ RN-AdS black holes \cite{Kubiznak:2012wp}.
	
\begin{figure}[htbp]
		\centering
		\includegraphics[height=50mm]{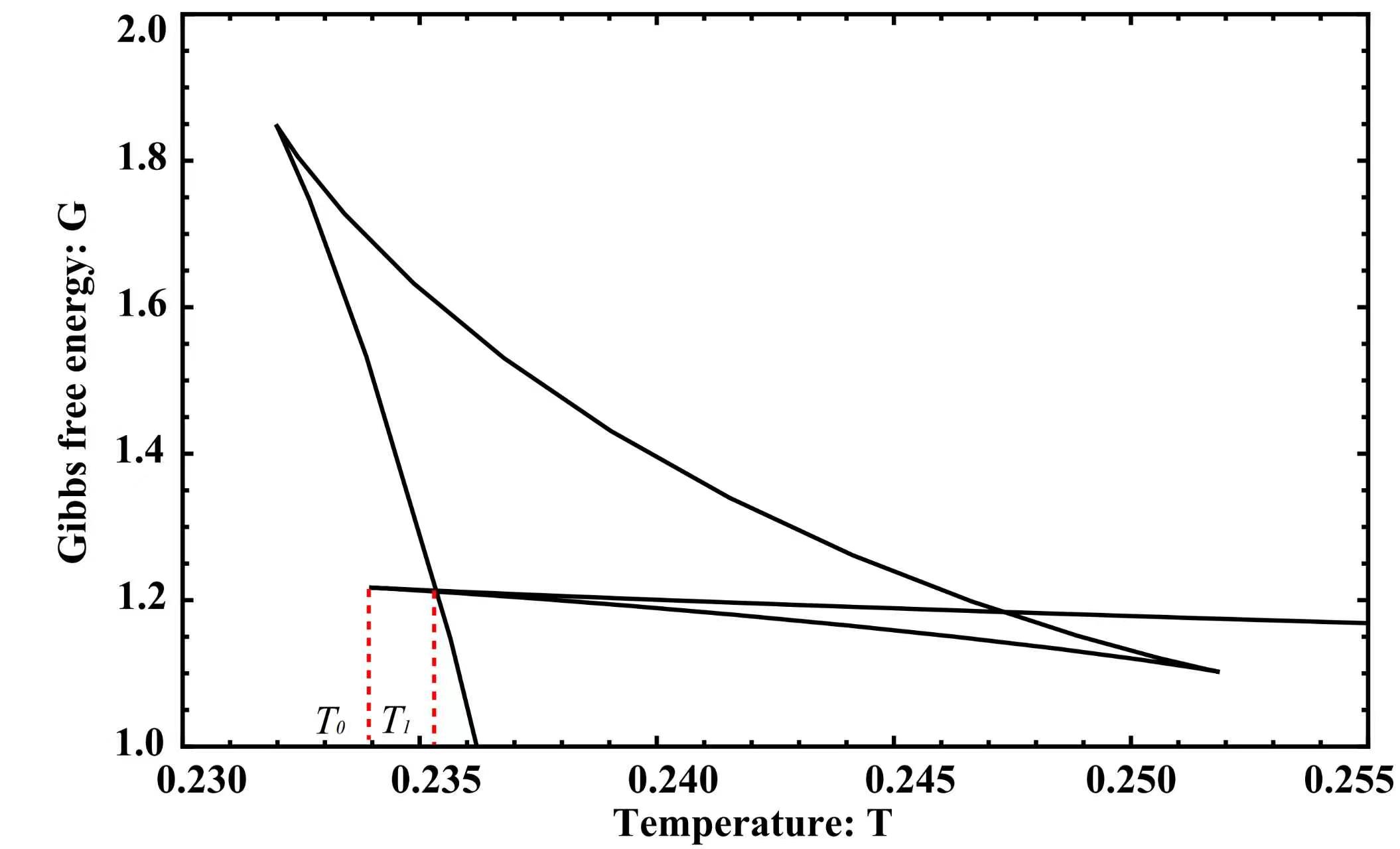}\hfill
		\includegraphics[height=50mm]{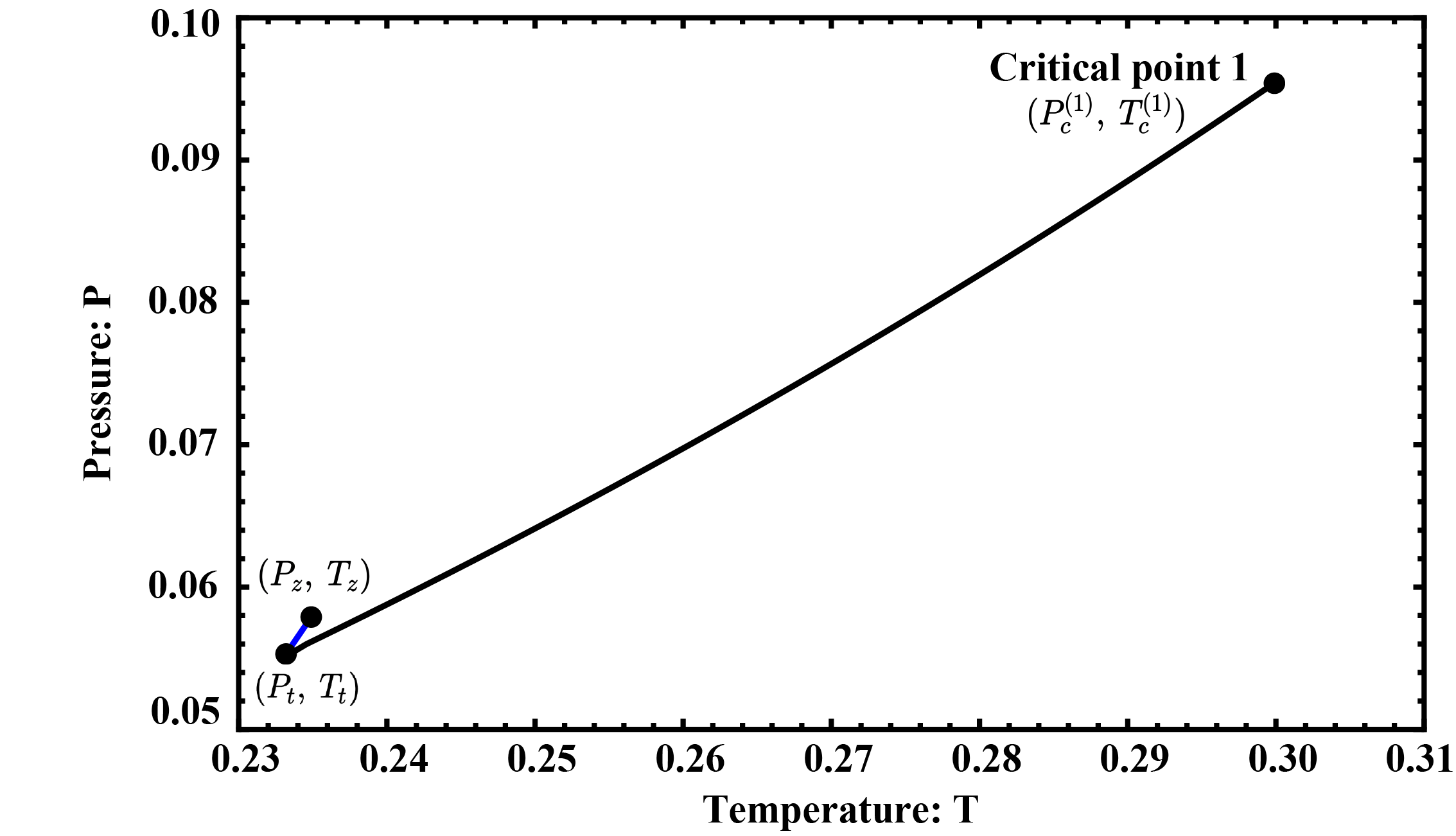}
		\caption{Phase diagram of Kerr-AdS black hole with $d=6$ and in the left diagram, we set $J=1$ and the pressure $P=0.0564$.}
		\label{fig2_aligned}
	\end{figure}

For $d \ge 6$, the system has two critical points and exhibits reentrant phase transition phenomena, which is shown in the left diagram in Fig.~\ref{fig2_aligned}, when the pressure is fixed at $P = 0.0564$. As the temperature decreases to $T_1$, the system undergoes a first-order phase transition from large black hole to small black hole. The Gibbs free energy of the two phases is continuous, but its first derivative is discontinuous, corresponding to the characteristics of a standard first-order phase transition. This phase transition process is depicted by a black line in the right diagram in Fig.~\ref{fig2_aligned}, showing a typical gas-liquid-type coexistence curve and ending at a critical point 1 $(P_{\rm c}^{(1)},T_{\rm c}^{(1)})$. When the temperature is further lowered to $T_0 < T_1$, the system undergoes a transition from small black hole to large black hole, with the Gibbs free energy exhibiting a discontinuous jump $\Delta G \neq 0$ at temperature $T_0$, marking the occurrence of a zeroth-order phase transition. This zeroth-order phase transition starts at the point $(P_z, T_z)$ and ends at the triple point $(P_t, T_t)$. This phase transition region is extremely narrow, and it is marked by the short blue solid line in the right diagram in Fig.~\ref{fig2_aligned}. It is worth noting that this blue line is not a traditional phase coexistence curve. Along this line, small black holes have a lower Gibbs free energy and remain in a locally stable state rather than coexisting phases; accordingly, the endpoint $(P_t, T_t)$ of this curve does not satisfy the critical point criterion. In a word, the first-order phase transition at high temperatures and the zeroth-order phase transition at low temperatures together form a complete reentrant phase diagram.
	
The above analysis establishes the fundamental thermodynamic framework for $d$-dimensional rotating AdS black holes and classifies the phase transition behavior below the critical point. In the following sections, we analytically continue the black hole horizon radius $r_h$ into the complex plane $z$. By applying the complex analysis framework proposed in~\cite{Xu2024}, we establish a physical correspondence between thermodynamic phase transitions and the Riemann surface structure of the analytic function. Furthermore, by incorporating the Lee-Yang theory of phase transitions, we enable an analytical treatment of black hole thermodynamic phase transitions. In this framework, the projection of Lee-Yang zeros onto the real phase space naturally gives rise to the supercritical behavior inherent in black hole thermodynamics~\cite{Xu:2025jrk}.

\section{Complex Analysis in the region below the critical point}\label{sec:subcritical_analysis}
	
In a recent study~\cite{Xu2024},  the generalized free energy based on the Maxwell equal-area law
\begin{equation}
U = \int (T - T_e) \mathrm{d}S,
\end{equation}
was constructed,
where $T_e$ is the canonical ensemble temperature and is just a positive constant which can be assigned in any way and $U, T, S$ are the functions of the event horizon $r_h$. When the integration is performed over the range from $S_1$ (entropy of the liquid-phase) to $S_2$ (entropy of the gas-phase), this construction reduces to the standard Maxwell equal-area law, implying the generalized free energy $U(r_h) \equiv 0$. For a non-vanishing generalized free energy $U(r_h)$, its extrema exactly satisfy the condition $T_e = T(r_h)$, implying that the system resides in a black hole state. It should be noted that the condition $T_e = T(r_h)$ is defined on-shell, and therefore cannot be directly incorporated into the definition of the generalized free energy. Furthermore, the characteristics of these extrema are intrinsically linked to the thermal stability of the system. Specifically, a convex extremum corresponds to a thermally unstable state, whereas a concave extremum indicates a thermally stable state. Naturally, thermal stability encompasses two regimes---metastable and stable states---which are determined by the local and global minima of the potential, respectively.

Mathematically, we associate the various states of the black hole thermodynamic system with the extrema of the generalized free energy. Consequently, the diverse thermal behaviors of the system can be characterized as the following function
\begin{equation}\label{eq:14}
\psi(r_h) := \frac{dU(r_h)}{dS(r_h)} = 0.
\end{equation}

To fully characterize the distribution of the zeros of $\psi(r_h)$, we extend the analysis into the complex plane, which enables a clear examination of their global behavior and topological properties. In other words, we analytically continue the functions $U(r_h)$ and $\psi(r_h)$  to  $U(z)$ and $\psi(z)$   via $r_h \rightarrow z := r_h + iy$ where $y$ is an arbitrary parameter. In complex analysis, the argument principle establishes a fundamental relationship between the contour integral of a meromorphic function $\psi(z)$ and the net number of its zeros and poles enclosed by the contour $C$. Specifically, for a meromorphic function $\psi(z)$ defined on and within a simple closed contour $C$, with no zeros or poles residing on $C$, the principle states that
\begin{equation}\label{eq:N-P}
N(\psi, C)- P(\psi, C) = \frac{1}{2\pi i} \oint_{C} \frac{\psi'(z)}{\psi(z)} dz = \frac{\Delta_C \arg \psi(z)}{2\pi},
\end{equation}
where $N(\psi, C)$ and $P(\psi, C)$ denote the number of zeros and poles of $\psi(z)$ enclosed by $C$, respectively, counted with their multiplicities,
with  $\psi'(z) = \frac{d\psi}{dz}$. The term $\Delta_C \arg \psi(z)$ signifies the total variation in the argument of $\psi(z)$ along the contour $C$, where $\psi(z) = |\psi(z)| \exp(i \arg \psi(z))$. Geometrically, eq.~(\ref{eq:N-P}) characterizes the winding number of the image curve $\bar{C}$ around the origin in the complex $w$-plane as $z$ traverses the contour $C$ under the transformation $w = \psi(z)$. Hence the winding number of $\bar{C}$ about the origin is
\begin{equation}
W := \frac{1}{2\pi i} \oint_{\bar{C}} \frac{\mathrm{d}\omega}{\omega} = \frac{1}{2\pi i} \oint_{C} \frac{\psi'(z)}{\psi(z)} \mathrm{d}z.
\end{equation}

Specifically, in the absence of poles for $\psi(z)$ within $C$, i.e., $P(\psi, C) = 0$, the winding number reduces to $W = N(\psi, C)$. As the complex variable $z$ traverses the contour $C$, the argument $\theta = \arg \psi(z)$ may exhibit multi-valued behavior. To restore single-valuedness, one typically constructs the corresponding Riemann surface. A complex function defined on each foliation (sheet) of the Riemann surface is inherently single-valued. Consequently, the winding number has a direct geometric interpretation: a winding number of $W$ corresponds to a Riemann surface consisting of $W$ foliations. In essence, the winding number provides a measure of the number of sheets of the Riemann surface associated with the argument function.

Next, we focus on exploring the phase transition characteristics of $d=4,5,6$-dimensional singly spinning Kerr-AdS black hole thermodynamics using complex analysis methods and establish some universal properties of phase transitions.

\subsection{4d Kerr-AdS Black Hole}\label{subsec:sub_4d_kerr}
In this case, the system undergoes a standard first-order phase transition, and its complete phase diagram is presented in Fig.~\ref{fig1_aligned}. Following the complex analysis method described above, the analyticfunction of this system can be directly obtained as
\begin{equation}\label{psi4}
\psi(z) =  \frac{3 z^4 + l^2 z^2 (1 - 4 \pi T_e z) + a^2 \left[ z^2 - l^2 (1 + 4 \pi T_e z) \right]}{ l^2 z (a^2 + z^2)},
\end{equation}
where $l = \sqrt{3/(8\pi P)}$.
 For fixed values of $T_e$ and $P$, the rule of signs indicates that the function possesses at most three real and positive zeros in the complex plane $\mathbb{C}\setminus\{0\}$; the condition  $J=1$ is imposed by setting Eq.~(\ref{eq:J})=1.
Hence the local maximum winding number is $W_{\text{local-max}} = 3$, and its complex counterpart is the Riemann surface with three foliations. For the Kerr AdS black hole, its thermodynamic behavior is similar to that of a Van der Waals fluid with a first-order phase transition.  We therefore see that a thermodynamic system will undergo a first-order phase transition, and the corresponding winding number is at least three. This observation is consistent with the ``Phase transition/Riemann surface'' correspondence~\cite{Xu2024}. While for the global winding number, we have $W_{\text{global}} =N(\psi, C)- P(\psi, C) = 4 - 3 = 1$ in the entire complex plane.
	
\subsection{5d Kerr-AdS Black Hole}\label{subsec:sub_5d_kerr}
	
In this case, we directly obtain
\begin{equation}\label{psi5}
\psi(z) = \frac{2 z^3 + a^2 (z - 2 \pi l^2 T_e) + l^2 z (1 - 2 \pi T_e z)}{ l^2 (a^2 + z^2)},
\end{equation}
where $l = \sqrt{3/(4\pi P)}$. The function possesses at most three real and positive zeros in the complex plane $\mathbb{C}\setminus\{0\}$. Hence the local maximum winding number is $W_{\text{local-max}} = 3$, and its complex counterpart is the Riemann surface with three foliations. Meanwhile for the global winding number, we have $W_{\text{global}} =N(\psi, C)- P(\psi, C) = 3 - 2 = 1$ in the entire complex plane.

\subsection{6d Kerr-AdS Black Hole}\label{subsec:sub_6d_kerr}
	
For the 6d Kerr-AdS black hole, the system undergoes the reentrant phase transition shown in Fig.~\ref{fig2_aligned}.  In this case
\begin{equation}\label{psi6}
\psi(z) =\frac{5z^4 + l^2z^2(3 - 4\pi zT_e) + a^2(3z^2 + l^2(1 - 4\pi zT_e))}{ l^2 z(a^2 + z^2)},
\end{equation}
where $l = \sqrt{5/(4\pi P)}$. In the complex plane $\mathbb{C}\setminus\{0\}$, $\psi(z)$ possesses at most four real and positive zeros. Consequently, its complex counterpart is treated as a Riemann surface with four foliations. While for the global winding number, we have   $W_{\text{global}} =N(\psi, C)- P(\psi, C) = 4 - 3 = 1$  in the entire complex plane. Based on the ``Phase transition/Riemann surface'' correspondence~\cite{Xu2024,Wang:2023qxw}, we see  that $W = 4$ can be decomposed in two distinct ways.
\begin{itemize}
\item[(i)] $ 4 = 2 + 2$: the system undergoes only two second-order phase transitions, corresponding to the critical point 1 $(P_{\rm c}^{(1)},T_{\rm c}^{(1)})$ and critical point 2 $(P_{\rm c}^{(2)},T_{\rm c}^{(2)})$ (the intersection of the orange and purple spinodal lines is shown in Fig.~\ref{Fig7} in the following text, where the critical point 2 is excluded from the phase diagram due to its higher Gibbs free energy compared to that of the extremal black hole).
\item[(ii)] $ 4 = 1 + 3$: the system has one first-order and one second-order phase transition.
\end{itemize}

From this, it can be observed that the zeroth-order phase transition neither contributes to the winding number nor alters the topological structure of the black hole thermodynamic system.
	
\section{Complex Analysis in the Supercritical Region}\label{sec:supercritical_analysis}
The core idea of the Lee-Yang phase transition theory is that the essence of a phase transition originates from the distribution of zeros of the partition function, namely the Lee-Yang zeros. In the thermodynamic limit, when these zeros approach the real axis and form a continuous distribution on it, the derivatives of thermodynamic functions (such as free energy) become non-analytic, and the system undergoes a genuine phase transition. In black hole thermodynamics, the Gibbs free energy $G$ of a black hole can be obtained by calculating the Euclidean action $I_E$, with the relationship between them given by $G=T I_E$. The thermodynamic partition function under the saddle-point approximation can be expressed as $Z = e^{-I_E}$, from which we can derive $G = -T \ln Z$. Hence, it is clear that the zeros of the partition function correspond to the singularities of the Gibbs free energy function (excluding extreme black holes) \cite{Xu:2025jrk}.

For the thermodynamics of black holes, the singularity of Gibbs free energy can be reflected in the thermodynamic response function, namely the singularity of isobaric heat capacity. Hence, the Lee-Yang zeros of the Kerr-AdS black hole thermodynamic system is determined by the following expression
	\begin{equation}
		C_{P, J} = T \left( \frac{dS}{dT} \right)_{P,J} \rightarrow \infty \quad \Rightarrow \quad
		\frac{\partial (T,J)}{\partial(z,a)} =
		\begin{vmatrix}
			\frac{\partial T}{\partial z} & \frac{\partial T}{\partial a} \\
			\frac{\partial J}{\partial z} & \frac{\partial J}{\partial a}
		\end{vmatrix}
		= \frac{\partial T}{\partial z} \frac{\partial J}{\partial a} - \frac{\partial T}{\partial a} \frac{\partial J}{\partial z} = 0.
		\label{eq:heat_capacity_divergence}
	\end{equation}
	
	\subsection{4D Kerr-AdS Black Hole}\label{subsec:super_4d}
\begin{figure}[htbp]
		\centering
		\includegraphics[width=120mm]{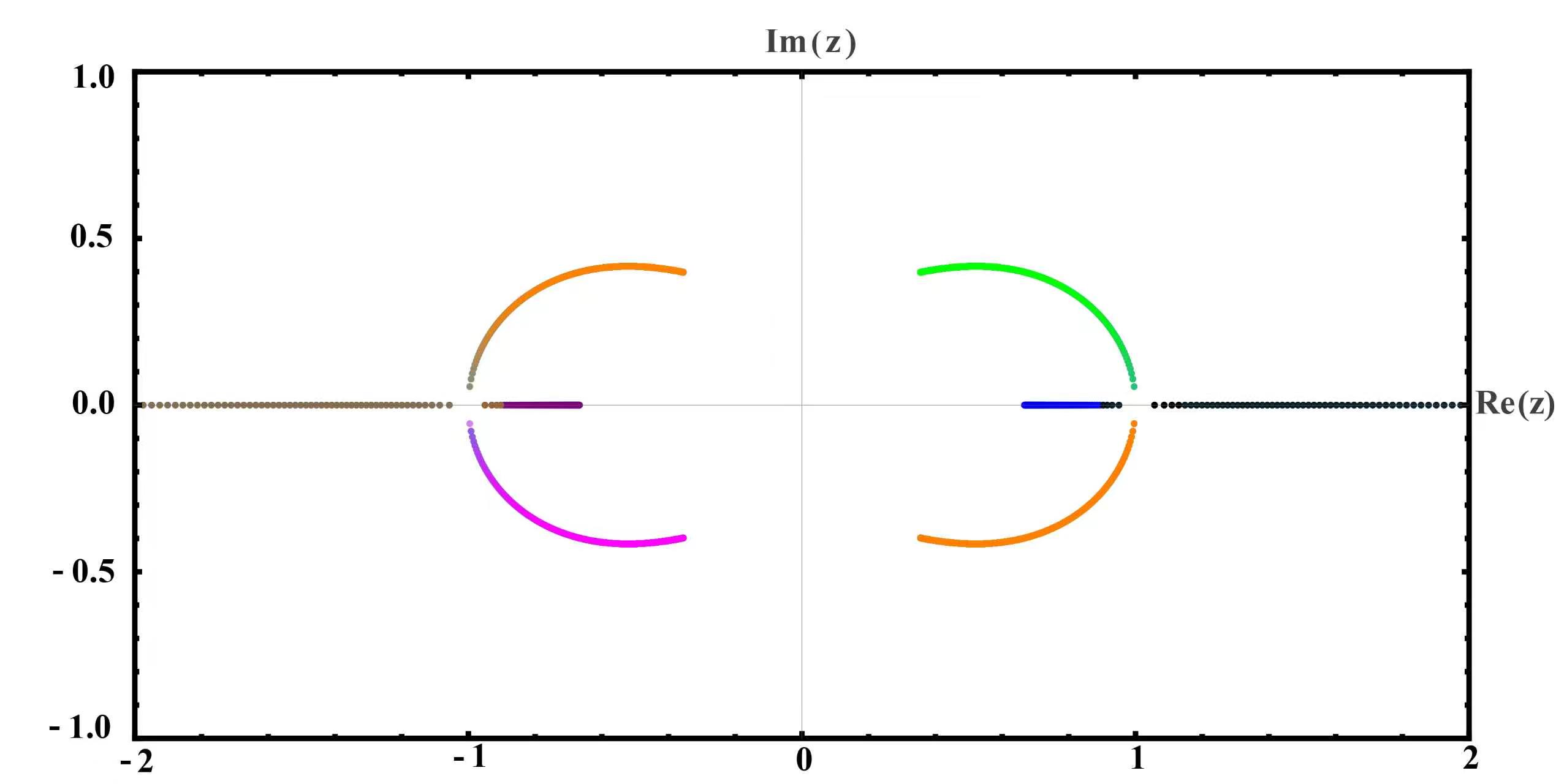}
		\caption{Singularity distribution of the Gibbs free energy for the Kerr-AdS black hole at $d=4$.}

		\label{Fig2}
	\end{figure}

\begin{figure}[htbp]
		\centering
		\includegraphics[width=170mm]{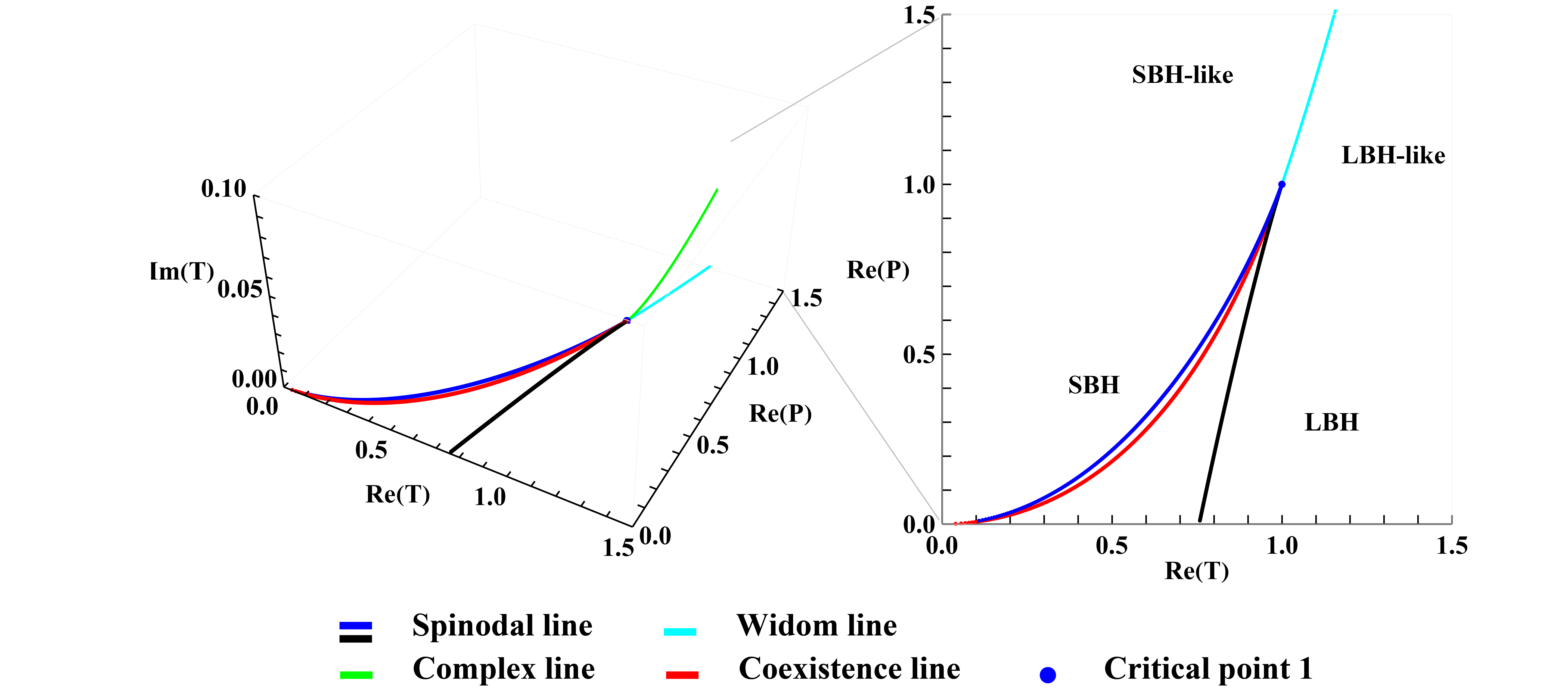}
		\caption{Complex phase diagram of Kerr-AdS black hole at $d=4$.}
		\label{Fig3}
	\end{figure}

	When the dimension $d=4$, the system undergoes the standard first-order phase transition shown in Fig.~\ref{fig1_aligned}. In this case, according to Eq.~(\ref{eq:heat_capacity_divergence}), we can obtain the distribution of the Lee-Yang zeros in Fig.~\ref{Fig2}, where we can see that zeros are divided into two categories: (i) when $P < P_c$, the zeros are on the real axis and the phase transitions in the subcritical region, or  (ii) $P > P_c$,   the zeros are distributed in the complex plane and the phase transitions are in the supercritical region. Correspondingly, the complex phase diagram for the black hole system is depicted in Fig.~\ref{Fig3}. From the two figures, we can obtion the information. (1) The coexistence line corresponding to the red line in the right diagram in Fig.~\ref{fig1_aligned} is obtained based on Maxwell's equal-area law~\cite{Spallucci2013}. It terminates at the critical point 1 $( P_{\rm c}^{(1)},T_{\rm c}^{(1)})$. Traversing this line corresponds to a first order phase transition from a large black hole to a small black hole.  (2) The spinodal lines (blue and black lines) correspond to the distribution of zeros with the same color in Fig.~\ref{Fig2} for $P < P_c$. These spinodal lines are the boundaries of the metastable states. (3) The complex line (green line) corresponds to the distribution of zeros with the same color in Fig.~\ref{Fig2}. It extends from the critical point 1 $(P_{\rm c}^{(1)},T_{\rm c}^{(1)})$ into the supercritical region. (4) The Widom line is the projection of the complex line onto the real plane. It is the boundary between two phases with different physical properties in the supercritical region, according to the idea presented in~\cite{Xu:2025jrk}.
		
\subsection{5D Kerr-AdS Black Hole}\label{subsec:super_5d}
\begin{figure}[htbp]
		\centering
		\includegraphics[width=120mm]{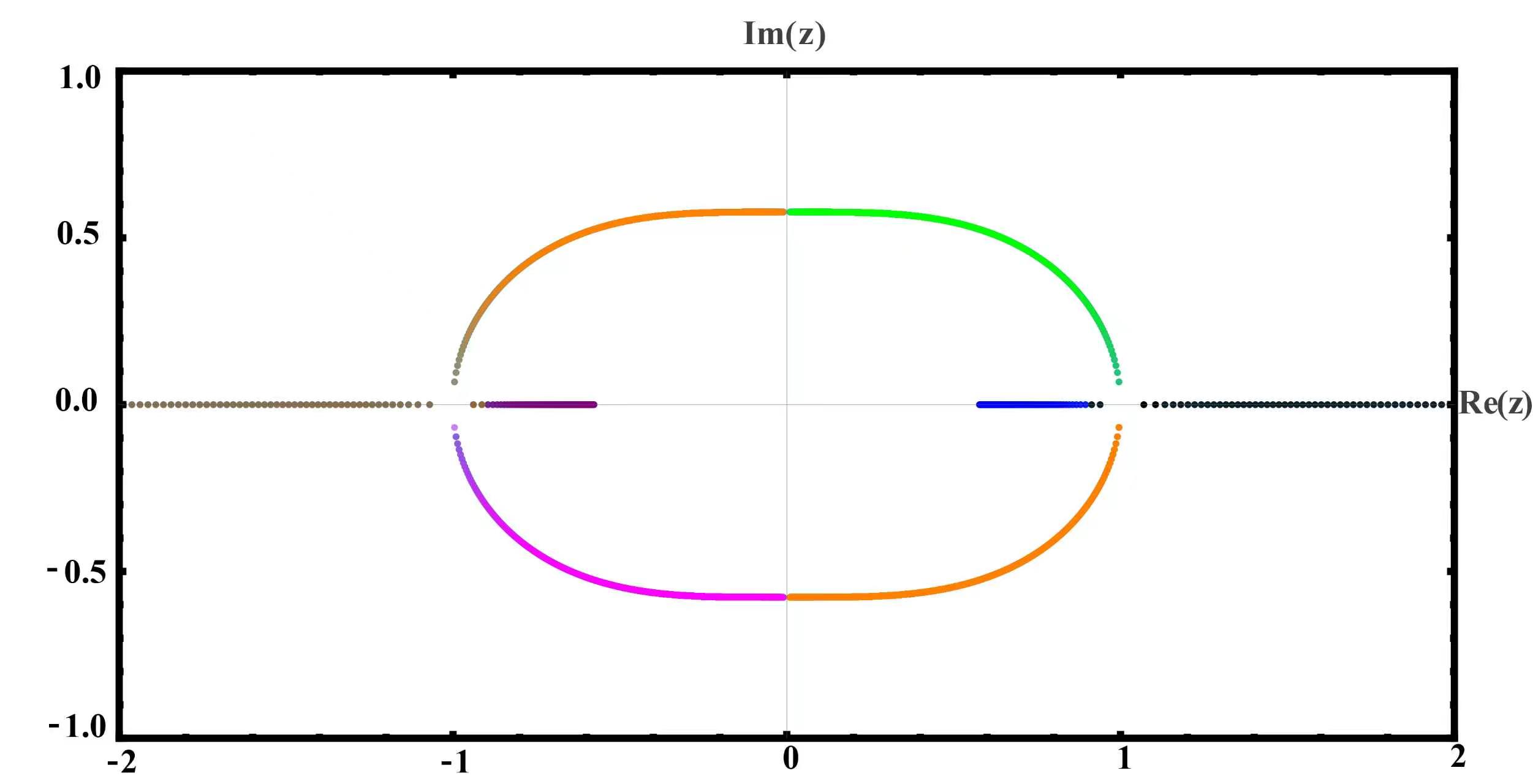}
		\caption{Singularity distribution of the Gibbs free energy for the Kerr-AdS black hole at $d=5$.}
		\label{Fig4}
	\end{figure}

\begin{figure}[htbp]
		\centering
		\includegraphics[width=170mm]{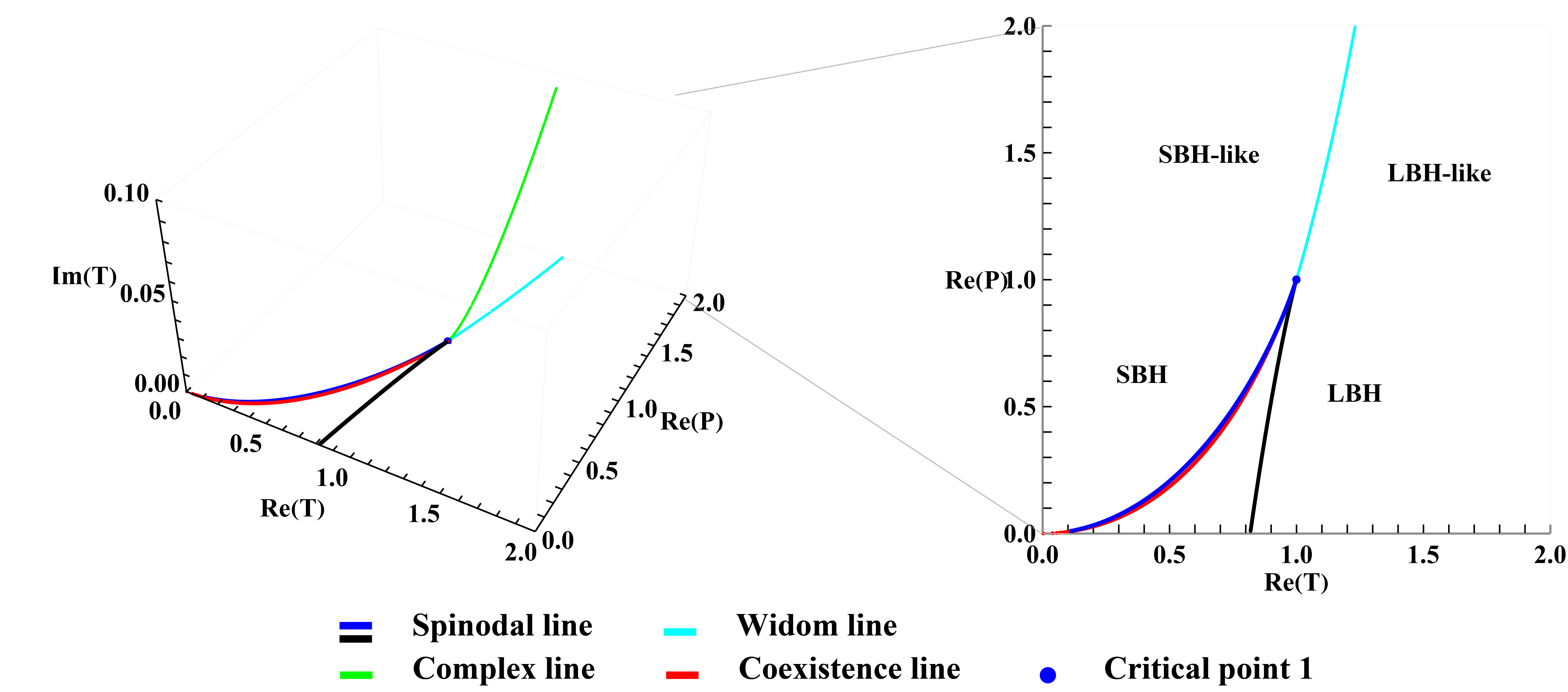}
		\caption{Complex phase diagram of Kerr-AdS black hole at $d=5$.}
		\label{Fig5}
	\end{figure}

From~(\ref{eq:heat_capacity_divergence}), we   obtain the distribution of Lee-Yang zeros shown  in Fig.~\ref{Fig4}. Correspondingly, the complex phase diagram for the black hole system is depicted in Fig.~\ref{Fig5}. Comparing Figs.~\ref{Fig3} and~\ref{Fig5}, we observe that the thermodynamic behaviors for four-dimension and five-dimension Kerr-AdS black hole are qualitatively identical.	
	
\subsection{6D Kerr-AdS Black Hole}\label{subsec:super_6d}

\begin{figure}[htbp]
		\centering
		\includegraphics[width=120mm]{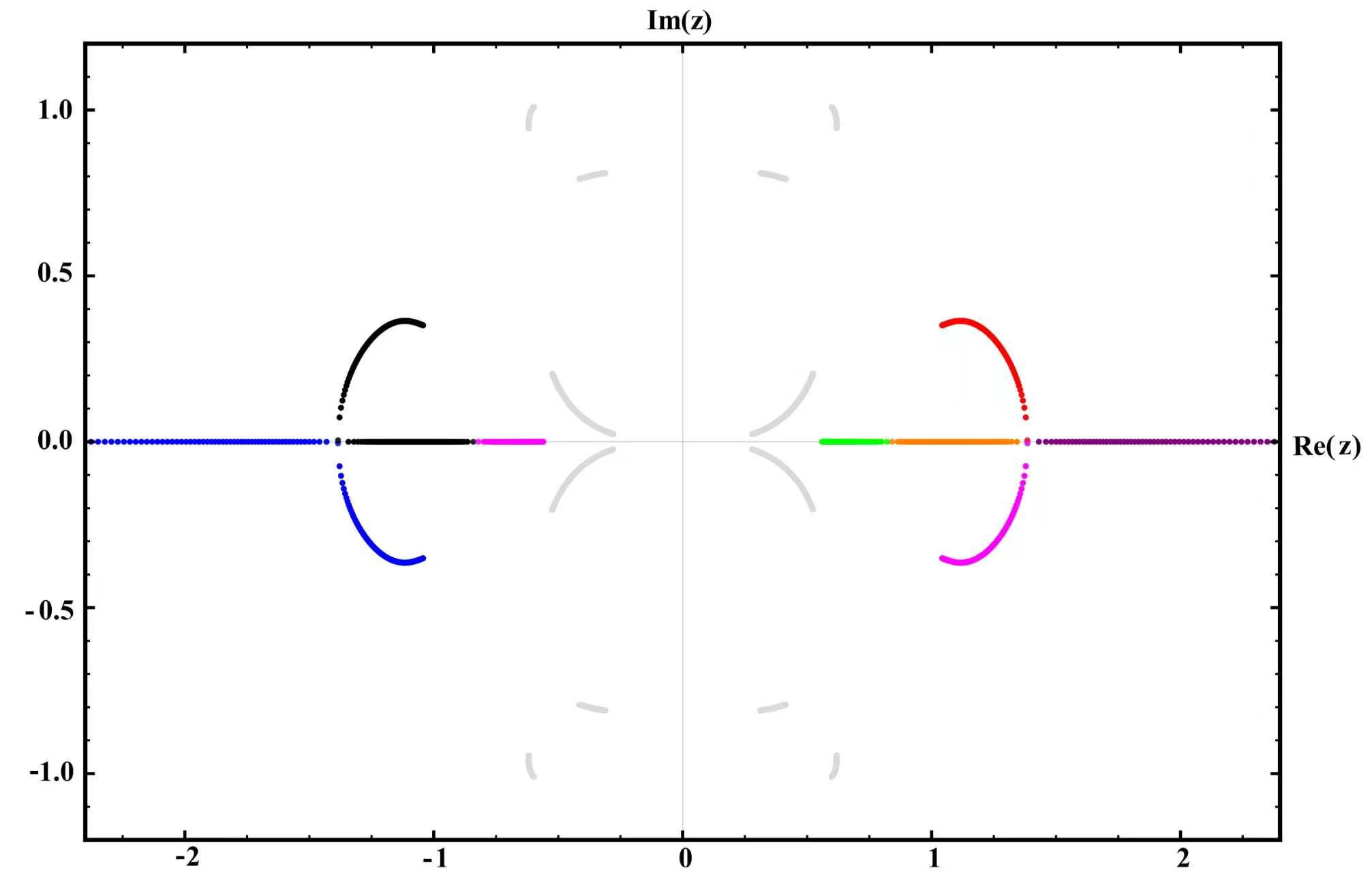}
		\caption{Singularities distribution of the Gibbs free energy for the Kerr-AdS black hole at $d=6$ (the gray markers represent unphysical solutions).}
		\label{Fig6}
	\end{figure}

When the dimension $d=6$, the system exhibits a reentrant phase transition, as shown in Fig.~\ref{fig2_aligned}. In this case, from ~(\ref{eq:heat_capacity_divergence}), we   obtain the distribution of Lee-Yang zeros in Fig.~\ref{Fig6}, where we can see that zeros are divided into two categories: (i) when $P < P_t$,   the zeros are on the real axis corresponding to phase transitions in the critical region, while (ii) $P > P_t$, o the zeros are distributed in the complex plane corresponding  to ``phase transitions'' in the supercritical region. Note that the physically meaningful zeros correspond to those in the first quadrant and on the positive real axis, where both the real and imaginary parts of the event horizon radius are positive.
	
\begin{figure}[htbp]
		\centering
		\includegraphics[width=170mm]{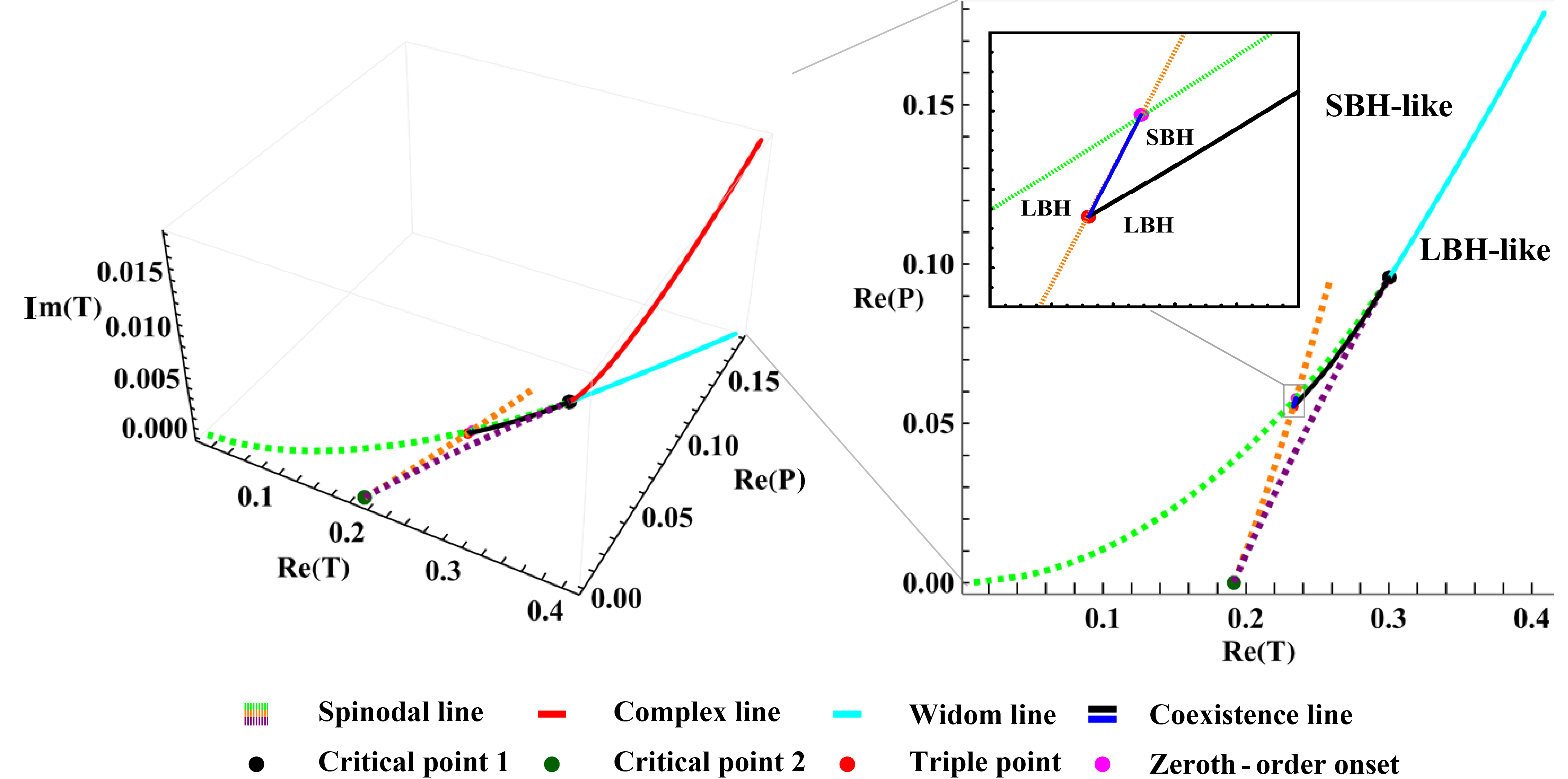}
		\caption{Complex phase diagram of Kerr-AdS black hole at $d=6$.}
		\label{Fig7}
	\end{figure}

Correspondingly, the complex phase diagram for the black hole system is depicted in Fig.~\ref{Fig7}. From this, we can obtain the following information.
	
	\begin{itemize}
		\item The coexistence line (the black line in Fig.~\ref{Fig7}, corresponding to the black line in the right diagram in Fig.~\ref{fig2_aligned}) is obtained through numerical calculations based on Maxwell's equal-area law.
		It originates from the critical point 1 $(P_{\rm c}^{(1)}, T_{\rm c}^{(1)})$ and terminates at the triple point ($P_{t}$, $T_{t}$). Traversing this line induces a phase transition from large black hole to a small black hole, which is a first-order phase transition. Note that this phase transition process is accompanied by an additional zero-order phase transition. Similarly, we can also obtain the zeroth-order phase transition ``coexistence line'' (the blue line in the right diagram in Fig.~\ref{Fig7},  corresponding to the blue line in the right diagram in Fig.~\ref{fig2_aligned}). It commences at the point ($P_{z}$, $T_{z}$) and also terminates at the triple point ($P_{t}$, $T_{t}$). Traversing this line corresponds to a phase transition from a small black hole to a large black hole. Both types of transitions are characterized by distinct discontinuities in thermodynamic state functions.
		\item The spinodal lines (green, purple and orange lines) correspond to the distribution of zeros for $P < P_t$: the orange and purple spinodal lines intersect at critical point 2 $(P_{\rm c}^{(2)}, T_{\rm c}^{(2)})$.  This critical point occurs at zero pressure and is therefore not included in the finite-positive-pressure phase diagram~\cite{Wei2016};
		 the intersection of the green and orange spinodal lines corresponds to the initiation point of the zeroth-order phase transition ($P_{z}$, $T_{z}$); and the green and purple spinodal lines intersect at critical point 1 $(P_{\rm c}^{(1)}, T_{\rm c}^{(1)})$. These spinodal lines are the boundaries of the metastable states.
		\item The complex line (red line) corresponds to the distribution of zeros in the complex plane. It extends from the critical point 1 $( P_{\rm c}^{(1)}, T_{\rm c}^{(1)})$ into the supercritical region. This line reflects the phase transition in the supercritical region.
		\item The Widom line is the projection of the complex line onto the real plane~\cite{Xu:2025jrk}. It marks the crossover line of the phase transition between large-like and small-like black holes in the supercritical region.
	\end{itemize}
	
	Based on the numerical calculations and analyses presented above, for Kerr-AdS black holes, we obtain a single Widom line in the supercritical region. This line originates from the stable critical point  1 $(P_{\rm c}^{(1)},T_{\rm c}^{(1)})$ of the first-order phase transition and extends into the supercritical region. Although the system exhibits a transition from standard first-order phase behavior to the reentrant phase transition phenomena in the region below the critical point as the dimension varies, its supercritical region maintains a single Widom line structure. Note that, as shown in Sec.~\ref{sec:thermo_found}, the Gibbs free energy is discontinuous across a zeroth-order phase transition. Therefore, no phase-transition critical point can be defined in the usual sense, and no corresponding Widom line structure arises beyond critical point $(P_z,T_z)$.

\section{Conclusion and Discussion}

In this paper, we developed a complex analytic approach to study thermodynamic phase transitions of singly rotating Kerr-AdS black holes. Below the critical point, we formulate the phase transition conditions in terms of the zeros of an analytic function in the complex plane. The number of these zeros is related to the local winding number. This relation connects the phase transition of the black hole with the complex structure for the analytic function and once again verifies the ``Phase transition/Riemann surface'' correspondence~\cite{Xu2024}. In the supercritical region, we employ the Lee-Yang theory to relate the singularities of the Gibbs free energy to the Lee-Yang zeros in the complex plane. The Widom line is then determined from the projection of the complex zeros onto the real parameter space.

The occurrence of a zeroth-order phase transition does not change the corresponding local winding number and also does not produce an additional Widom line in the supercritical region, although it changes the thermodynamic phase structure below the critical point. A remaining question is how the occurrence of a zeroth-order phase transition is encoded in the complex zeros. In particular, it is important to determine whether the complex zeros show a characteristic evolution at such a transition and whether this evolution can provide a criterion for identifying zeroth-order phase transitions. These topics require further in-depth analysis in the future.

\section*{Acknowledgments}
This research was supported in part by the National Natural Science Foundation of China (Grant Nos. 12575064, 12247103), by the Natural Science Basic Research Plan in Shaanxi Province of China (Grant No. 2025JC-YBQN-029),   by the project of Tang Scholar in Northwest University, and by the Natural Sciences and Engineering Research Council of Canada.

\end{document}